\documentclass[fleqn,usenatbib]{mnras}

\pdfoutput=1 

\usepackage{newtxtext,newtxmath}
\usepackage[T1]{fontenc}
\usepackage{ae,aecompl}
\usepackage{hyperref} 
\usepackage{verbatim} 
\usepackage{graphicx}
\usepackage{subcaption}
\usepackage[table,usenames,dvipsnames]{xcolor}

\def\be{\begin{equation}} 
\def\ee{\end{equation}} 
\def\bea{\begin{eqnarray}} 
\def\eea{\end{eqnarray}} 

\title[Cosmic string locations on CMB maps]{Inferring Cosmic String Tension through the Neural Network Prediction of String Locations in CMB Maps}{}

\author[R. Ciuca and O. F. Hern\'andez]{
Razvan Ciuca$^{1,2,3}$\thanks{Email: razvan.ciuca@mail.mcgill.ca}
Oscar F. Hern\'andez$^{1,2}$\thanks{Email: oscarh@physics.mcgill.ca}
\\
$^{1}$Department of Physics, McGill University, 3600 rue University, Montr\'eal, QC, H3A 2T8, Canada
\\
$^{2}$Marianopolis College,  4873 Westmount Ave.,Westmount, QC H3Y 1X9, Canada
\\
$^{3}$
School of Computer Science, McGill University,  3480 rue University, Montr\'eal, QC, H3A 0E9, Canada
}

\date{}
\pubyear{2018}

\begin{document}
\label{firstpage}
\pagerange{\pageref{firstpage}--\pageref{lastpage}}
\maketitle

\begin{abstract}
In previous work, 
we constructed a convolutional neural network used to estimate the location of cosmic strings in simulated cosmic microwave background temperature anisotropy maps. We derived a connection between the estimates of cosmic string locations by this neural network and the posterior probability distribution of the cosmic string tension $G\mu$. 
Here, we significantly improve the calculation of the posterior distribution of the string tension $G\mu$. We also  improve our previous plain convolutional neural network by using residual networks. We apply our new neural network and posterior calculation method to maps from the same simulation used in our previous work 
and quantify the improvement. 
\end{abstract}


\section{Introduction}
\label{sec:intro}
In recent years there has been a renewed interest in cosmic strings since they can form in a large class of extensions of the Standard Model.  Cosmic strings are linear topological defects, remnants of a high-energy phase transition in the very early Universe.  
The gravitational effects of the string can be parametrized by its string tension $G\mu$, a dimensionless constant where $G$ is Newton's gravitational constant, and $\mu$ is the energy per unit length of the string. 
Because of the continued disagreement between Nambu-Goto simulations~\citep{Ringeval:2007gf, Lorenz:2010iq,BlancoPillado:2014kr} and Abelian Higgs simulations~\citep{Hindmarsh:2017iw} regarding the distribution of cosmic string loops in a network, the robust limit on the string tension is provided by long string effects. The best limits come from the Planck collaboration's analysis of the long string contribution to the angular power spectrum: $G\mu<1.3\times10^{-7}$ and $G\mu<3.2\times10^{-7}$ at the 95\% confidence level (CL) for Nambu-Goto and Abelian Higgs strings, respectively~\citep{PlanckCollaboration:2014il}.  

Long strings moving between an observer and the surface of last scattering induce a step discontinuity in a cosmic microwave background (CMB) temperature anisotropy map through the Gott-Kaiser-Stebbins (GKS) effect~\citep{Gott:1985eg,Kaiser:1984jg}.  Searches for the GKS effect in the data from {\it Wilkinson Microwave Anisotropy Probe (WMAP)} and {\it Planck} have been considered as a way complementary to the angular power spectrum to detect strings. 
\cite{Lo:2005tm}, \cite{Jeong:2005bg,Jeong:2007jl}, and \cite{Jeong:2010dh} looked at WMAP data to search for this effect and this leads to null detection and a limit of $G\mu< 1.5 \times 10^{-6}$. The \cite{PlanckCollaboration:2014il} also did non-Gaussian searches for strings through the GKS effect and obtained constraints of $G\mu \lesssim 7.8 \times 10^{-7}$. 

In~\cite{Ciuca:2017jz} and~\cite{Ciuca:2017ww} we proposed a Bayesian interpretation of cosmic string detection where we developed a convolutional neural network to estimate the cosmic string locations in CMB maps by detecting the GKS effect.  Furthermore, we derived a connection between these location estimates and the posterior probability distribution of the cosmic string tension $G\mu$.  However our previous work and the work in this paper are not the first to consider the GKS effect to detect strings in simulations.  

A series of works has used a scale invariant analytic model of long straight strings described in~\cite{Perivolaropoulos:1993efa} to simulate CMB temperature anisotropy maps. These works then studied the limits that different algorithms could place 
on the string tension through the GKS effect in the simulated maps. In particular, \cite{Amsel:2008it}, \cite{Stewart:2009fr}, and \cite{2010IJMPD..19..183D} used the Canny algorithm~\citep{Canny:et} and found more short edges in maps with strings that they interpreted as the disruption of long edges by Gaussian noise. \cite{Stewart:2009fr} claimed a detection limit with noise of $G\mu=5.5\times10^{-8}$. In~\cite{Ciuca:2017jz} we reproduced the Canny algorithm analysis of those works and showed that the excess number of short edges found did not correspond to the locations of cosmic strings and that the limits of detection provided by Canny are for string tension above $G\mu \sim 10^{-6}$. \cite{Movahed:2011em} studied the simulated maps with a level crossing statistic and claimed that strings could be detected in noiseless maps when $G\mu>4\times10^{-9}$.  In another example \cite{Hergt:2017dr} used wavelets, and curvelets and found that strings could be detected down to a string tension of $G\mu=1.4 \times 10^{-7}$ at the 95\% CL if the contribution of noise was not more than $1.6\mu$K (see their table III). 

Another series of works has studied simulations of realistic Nambu-Goto strings in the full sky~\citep{Ringeval:2012gp} or the flat sky~\citep{Fraisse:2007nu}. 
\cite{McEwen:2017cg} considered ``Planck-like'' full sky realistic Nambu-Goto string simulations with noise and were able to recover accurate estimates of the string tension for simulation with $G\mu$ as low as $5\times10^{-7}$. These simulations included only the Integrated Sachs-Wolfe (ISW) effect and so the resulting string-tension sensitivity when considering them is necessarily more conservative than those works that rely only on the straight string GKS effect.  

\begin{figure}
\includegraphics[width=\columnwidth]{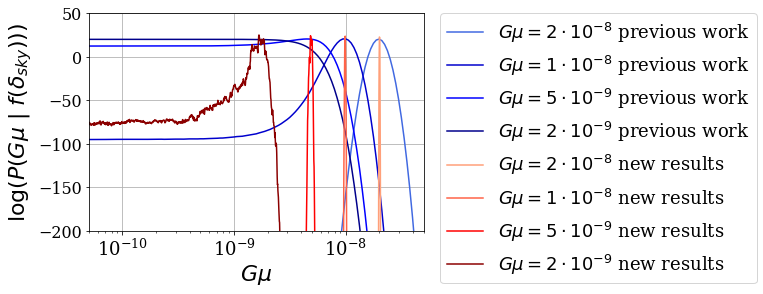}
\caption{Comparison of the string tension posteriors obtained in our previous work 
to the posteriors obtained after the improvements discussed in this paper. These improvements include an improved neural network and  the calculation of the posterior using equation~(\ref{Ppapprox}).}
\label{fig:v2paper_to_v6resnet}
\end{figure}

Most recently, \cite{VafaeiSadr:2018hh} used a pipeline of curvlets, the Canny algorithm, and other statistical tools on flat sky Nambu-Goto simulations and claimed that strings with tension as low as $1.2\times10^{-7}$ and $4.3\times10^{-10}$ could be detected on maps with and without noise, respectively. 
And \cite{VafaeiSadr:2018bc} used tree-based machine learning algorithms to claim detection limits of $3.0\times10^{-8}$ and $2.1\times10^{-10}$ in maps with and without noise. These detection limits are different than measurement limits as shown in \cite{VafaeiSadr:2018bc} tables 1 and 2. Their measurement limits are $1.2\times10^{-7}$ and $3.6\times10^{-9}$ for maps with noise and without noise, respectively. The $G\mu$ measurement limit from~\cite{Ciuca:2017jz} fig.~8 is $4\times10^{-9}$, which is comparable to their noiseless measurement limit. Also our work produces estimates of string locations, which increases the verifiability of our method. In this paper, we significantly improve on both of these points (Figure~\ref{fig:v2paper_to_v6resnet}).
This significant improvement includes two effects: an improved neural network and an improvement  in our calculation of the posterior probability of the string tension, in particular using equation~(\ref{Ppapprox}). 

The training of a neural networks requires much data. This is particularly true in our case, where we wish for the network to learn to identify string locations and not just produce a value for the string tension. Hence for our network to learn where the strings may be, we need a large amount of training maps. Whereas Nambu-Goto simulations are computationally costly, the straight string simulations of \cite{Perivolaropoulos:1993efa} are not.  In ~\cite{Ciuca:2017jz} and \cite{Ciuca:2017ww}, we used this model for our simulations. Furthermore in~\cite{Ciuca:2017ww} we showed that even though our network was developed and trained on the Perivolaropoulos model, it was still capable of accurate estimates of string locations in Nambu-Goto simulations with $G\mu=5\times10^{-8}$. This is further evidence 
that our network was learning about small-scale step discontinuities and not about some feature particular to the numerical simulation.  

One of the unknown parameters characterizing the scaling solution of strings is the number of strings per Hubble volume, $N_H$, which can have a value between 1 and 10. We trained our neural network on simulations with a value of $N_H=1$ and this did not impair the predictive power for input maps with larger $N_H$ values. This is an indication that the network is indeed generalizing and not just overfitting. We did our testing of the predictive power of the network with $N_H=3$. 

Our ultimate goal is to use our network on real flat sky CMB data such as that from the South Pole Telescope~\cite{Chown:2018ez} and the Atacama Cosmology Telescope~\cite{Louis:2014ef}. That is why we consider simulation of $512\times512$ pixels with 1 arcmin resolution per pixel. This corresponds to about a $8.5^\circ\times8.5^\circ$ patch of sky. Though CMB data are stored in \textsc{healpix} format, we worked with a square grid because the \textsc{pytorch} environment (pytorch.org) in which we develop our convolutional neural network has convolutions optimized at the graphics processing unit (GPU) instruction level for square grids.  

In \cite{Ciuca:2017jz}, we showed we could derive sharply peaked posteriors for string tensions as low as $4\times10^{-9}$ on noiseless maps. 
We modelled the CMB map  $\delta_{sky}$ as being composed of two different elements, $\delta_{gauss}$ and $\delta_{string}$, such that $\delta_{sky} = \delta_{gauss} + G\mu\,  \delta_{string}$.  The $\delta_{gauss}$ term is the standard $\Lambda$CDM cosmology CMB anisotropies that can be computed from the power spectrum, whereas $\delta_{string}$ is made up of the superposition of GKS temperature discontinuties of individual strings, each given by 
$
8 \pi G \mu \gamma_s [\hat{n}\cdot(\vec{v}_s\times\hat{e}_s)]
$,
where $\hat{n}$ is the direction of observation, $\vec{v}_s$ is the velocity of the string, $\hat{e}_s$ is its orientation, and  $\gamma_s = (1-v_s^2/c^2)^{-1/2}$.

To use our network on real data, we need to improve our network architecture so that it can produce sharply peaked string tension posteriors for Nambu-Goto simulation maps in \textsc{healpix} format with realistic noise and with $G\mu$ well below $10^{-7}$.  Only then can we confidently feed the network real data and provide new robust string tension limits, or a detection!  

A first modest step to achieving our ultimate goal is to improve and extend our previous network architecture and calculation of string tension posteriors to smaller string tensions. 
In this paper, we do just that. We present significant improvements to the calculation of the posterior distribution of the string tension $G\mu$ and to the convolutional neural network presented in~\cite{Ciuca:2017jz} and \cite{Ciuca:2017ww}. These improvements are summarized in Fig.~\ref{fig:v2paper_to_v6resnet} where we plot the logarithm of the posterior probability of the string tension versus string tension. 

The improvement in the calculation of the posterior probability is presented in Section~\ref{better_posterior} and it involves two points. The first is a reformulation of the posterior probability in terms of the evidence provided by the neural network evaluated on the sky map, rather than the sky map itself. This allows for a more general and more precise interpretation of our posterior probability formula, as discussed in subsections~\ref{better_posterior_f} and~\ref{better_posterior_conditions}. It also allows us to derive a more efficient and accurate way to compute the posterior probability as presented in subsection~\ref{more_efficient}. In subsection~\ref{oldnn}, we present the improved posteriors obtained by applying formula~\ref{Ppapprox} to the prediction maps using the neural network presented in~\cite{Ciuca:2017jz} and \cite{Ciuca:2017ww}. 

The improvement to the plain convolutional neural network involves the use of residual networks~\cite{He:2015tt}. The residual network and the results obtained with it and the new posterior calculation are presented in subsection~\ref{rn} and~\ref{rnpredictions}. We also compare the prediction maps and posteriors with the results obtained using the old network. We quantify the improvement of the prediction maps by using the standard deviation of the prediction values of the pixel. We quantify the information gained between the old and new posterior distributions by calculating the 
Kullback-Leibler (KL) divergence between them. 

\section{A better way to compute the posterior probability}
\label{better_posterior}
\subsection{The posterior probability in terms of the evidence provided by the neural network}
\label{better_posterior_f}
Equation (2.4) of reference~\cite{Ciuca:2017jz} expressed the posterior probability of the distribution of the string tension $G\mu$, $P(G\mu \, | \, \delta_{sky})$, given the sky map $\delta_{sky}$ as evidence. We present it here as equation~(\ref{PGsky}):
\begin{multline}
\label{PGsky}
P(G\mu \, | \, \delta_{sky})  \\ = 
\Big({1\over2}\Big)^{N_{\rm pixel}}\, 
\Bigg(
\frac{P(G\mu)}{P(\delta_{sky})} 
\Bigg)
\Bigg\{\sum_{\xi \in \mathbf{\Xi}}\frac{P(\delta_{sky} \, | \, \xi, G\mu) \times P(\xi)}{P(\xi \, | \, \delta_{sky}, G\mu)}\Bigg\}\, . 
\end{multline}
$\xi$ is a map which indicates which pixels lie on a string. If $(i,j) \in \textit{string}$ then $\xi_{i,j}=1$, otherwise $\xi_{i,j}=0$. A map $\xi$ is associated with a CMB temperature map $\delta_{sky}$.  We call the space of all such maps $\mathbf{\Xi}$ and it contains $2^{N_{\rm pixel}}$ elements where $N_{\rm pixel}$ is the number of pixels in a map. This formula uses information about the string locations to update our knowledge of the prior distribution $P(G\mu)$ to the posterior $P(G\mu \, | \,  \delta_{sky})$.

We can be more general, and more precise, by considering $f(\delta_{sky})$ instead of $\delta_{sky}$ as evidence, where  $f$ can be any function, but we are interested in the case where $f$ is a convolutional neural network. We treat the output of $f$ as evidence and compute the posterior with respect to it: $P(G\mu \, | \, f(\delta_{sky}))$. If $f$ is one-to-one, this is the same posterior as $P(G\mu \, | \, \delta_{sky})$. 

In Appendix~\ref{bayes_derivation} we derive the equivalent of equation~(\ref{PGsky}) for this case:
\begin{multline}
\label{PpGfsky_pre}
P(G\mu \, | \,f(\delta_{sky})) \; \\ =
\Big({1\over2}\Big)^{N_{\rm pixel}}\,  
\frac{P(G\mu)}{P(f(\delta_{sky}))} \; {1\over N} \sum_{\xi^a \sim P(\xi)} \frac{P( f(\delta_{sky})\, | \, \xi, G\mu) }{ P(\xi \, | \, f(\delta_{sky}), G\mu)} 
\end{multline}
where we have transformed the sum of $\xi \in \mathbf{\Xi}$ into an expectation of $N$ maps $\xi^a$ sampled from $P(\xi)$. Thus the factor $P(\xi)$ is not in the summand but in the sampling procedure. Also notice that $2^{N_{pix}}$, $P(f(\delta_{sky}))$, and $N$ are $G\mu$ independent, hence we can absorb them into the normalisation ($\int P(G\mu \, | \,f(\delta_{sky})) dG\mu = 1$). Thus  we only need to compute the unnormalized probability $P'(G\mu \, | \,f(\delta_{sky}))$:
\be
\label{PpGfsky}
P'(G\mu \, | \,f(\delta_{sky})) \; = P(G\mu) \; \sum_{\xi^a \sim P(\xi)} \frac{P( f(\delta_{sky})\, | \, \xi^a, G\mu)}{ P(\xi^a \, | \, f(\delta_{sky}), G\mu)}
\ee

\subsection{Conditions on the neural network ${\boldsymbol f}$}
\label{better_posterior_conditions}
We now make some assumptions regarding the denominator, $P(\xi \, | \, f(\delta_{sky}), G\mu)$, and the numerator, $P( f(\delta_{sky})\, | \, \xi, G\mu)$, of the summand of equation~(\ref{PpGfsky}). As in ~\cite{Ciuca:2017jz}, these assumptions encodes our conjecture that we should be able to decide whether a given pixel is on a string without knowing anything about which other pixels are actually on a string. Notice that in this abstraction $f$ is completely free, any function that respects our assumptions below can be used. Most functions would make the assumptions below quite bad, however, these assumptions are reasonable for the case of the function represented by our neural networks.

1. Conditional independence of each pixel ${i,j}$ in the answer map $\xi$ and $f(\delta_{sky})$:
\bea
P(\xi \, | \, f(\delta_{sky}), G\mu) = & \prod_{i,j} P(\xi_{i,j} \, | \, f(\delta_{sky}), G\mu)  
\nonumber \\ 
P(f(\delta_{sky}) \, | \, \xi, G\mu) = & \prod_{i,j}P(f_{i,j}(\delta_{sky}) \, | \, \xi, G\mu)   \nonumber
\eea

2.  Each pixel ${i,j}$ in the answer map depends only on the corresponding pixel in $f(\delta_{sky})$ and vice versa:
\bea
P(\xi_{i,j} \, | \, f(\delta_{sky}), G\mu) = P(\xi_{i,j} \, | \, f_{i,j}(\delta_{sky}), G\mu) 
\nonumber \\
P(f_{i,j}(\delta_{sky}) \, | \, \xi, G\mu) \, = \, P(f_{i,j}(\delta_{sky}) \, | \, \xi_{i,j}, G\mu) 
\nonumber
\eea

3. Translation invariance in the probabilities, i.e. pixel location is not important:
\bea
&&\forall\ i, j, i', j' ~{\rm where}~\xi_{i,j}=\xi_{i',j'}~{\rm and}~ f_{i,j}=f_{i',j'} \nonumber
\\
&&P(\xi_{i,j} \, | \, f_{i,j}(\delta_{sky}), G\mu) = P(\xi_{i',j'} \, | \, f_{i',j'}(\delta_{sky}), G\mu) \nonumber
\\
&&~~~~~~~~~~~~~~~~~~~~~~~~~~~{\rm and}
\nonumber\\
&&P(f_{i,j}(\delta_{sky}) \, | \, \xi_{i,j}, G\mu) = P(f_{i',j'}(\delta_{sky}) \, | \, \xi_{i',j'}, G\mu) \nonumber
\eea

At this point the only quantities we need to compute are the pixel-independent probabilities $P(\xi_{i,j} \, | \, f_{i,j}(\delta_{sky}), G\mu)$ and $P(f_{i,j}(\delta_{sky}) \, | \, \xi_{i,j}, G\mu)$. These are tractable quantities which we can easily compute from data.

To compute $P(\xi_{i,j} \, | \, f_{i,j}(\delta_{sky}), G\mu)$ we begin with a collection of simulated sky maps with strings of a known $G\mu$ and we bin the values that $f$ takes on these maps. For each specific bin, we then take the fraction of pixels which are on strings among the pixels with the specified value of $f$ and assign this value $p$ to $P(\xi_{i,j}=1 \, | \, f_{i,j}(\delta_{sky})$ and $1-p$ to $P(\xi_{i,j}=0 \, | \, f_{i,j}(\delta_{sky})$. 

To compute the probability distribution $P(f_{i,j}(\delta_{sky}) \, | \, \xi_{i,j}, G\mu)$ we again begin with a collection of simulated sky maps with strings of a known $G\mu$. We then consider all the pixels in our data set with $\xi_{i, j} = 1$ and compute the histogram of values of $f$ on those pixels. We do the same thing for those pixels with  $\xi_{i, j} = 0$.

For the calculations of $P(\xi_{i,j} \, | \, f_{i,j}(\delta_{sky}), G\mu)$ and $P(f_{i,j}(\delta_{sky}) \, | \, \xi_{i,j}, G\mu)$ that lead to the results we present in Section~\ref{nn} we binned the values of $G\mu$ in 700 bins of equally spaced log intervals between $10^{-11}$ and $2\times10^{-7}$:
$$G\mu=10^{-11+n\times(4+\log_{10}2) / 700} ~,~~n=0,..., 700~,
$$
and the values of $f$ in 1000 equally spaced bins between 0 and 1 with size 1/1000.

Now that we have the summand it remains to compute the sum over $\xi$ in equation~(\ref{PpGfsky}). 

\subsection{A more efficient way to compute the summand in equation~(\ref{PpGfsky})}
\label{more_efficient}

Consider the map $\xi^*$ that maximizes the summand 
\be
s(\xi^a)\equiv {P(f(\delta_{sky}) \, | \, \xi^a, G\mu) \over P(\xi^a \, | \, f(\delta_{sky}), G\mu)} \ .
\ee
$s(\xi^*)$ is then the largest term in the sum of equation~(\ref{PpGfsky}).
Note that $\xi^*$ depends on $G\mu$. With the distributional assumptions made in the last section, such a map is easily computable. Because the probability factorises over pixels we can find the maximal map by optimizing each pixel independently. In particular, since
\be
s(\xi)= \prod_{i,j} {P(f_{i,j}(\delta_{sky}) \, | \, \xi_{i,j}, G\mu)  \over
				P(\xi_{i,j} \, | \, f_{i,j}(\delta_{sky}), G\mu)}
	\equiv \prod_{i,j} s_{i,j}(\xi_{i,j})
\ee
we have that
\be
\xi^*_{i,j}  = 
\begin{cases} 
1, & \mbox{if } s_{i,j}(\text{\small$\xi_{i,j}=1$})~>~ s_{i,j}(\text{\small$\xi_{i,j}=0$})
\\ 
0, & \mbox{otherwise}  \end{cases}
\ee
Hence computing the maximal map from $f(\delta_{sky})$ is computationally straightforward. 

Now consider a map $\xi^*_{-1}$ identical to $\xi^*$ except at 1 pixel, at which the value of $\xi^*_{-1}$ is the opposite of the corresponding value of $\xi^*$. Since $\xi^*$ is by definition the map which maximizes the probability, the map with 1 pixel reversed will decrease the probability by some value, the new log probability is 
\begin{multline}
\log s(\xi^*_{-1}) 
=
\sum_{i,j} \log\bigl( s_{i,j}(\xi^*_{-1, i,j}) \bigr)
 \\
~~~~~~=
\log\bigl( s_{i',j'}( 1-\xi^*_{i',j'})  \bigr) + \sum_{i,j \neq i', j'} \log\bigl( s_{i,j}(\xi^*_{i,j})  \bigr) 
 \\
=
\log\bigl( s_{i',j'}( 1-\xi^*_{i',j'})  \bigr) - \log\bigl( s_{i',j'}( \xi^*_{i',j'})  \bigr) + \log\bigl( s(\xi^*)  \bigr)
\end{multline}
where $(i', j')$ is the pixel by which $\xi^*$ and $\xi^*_{-1}$ differ.

We can compute the expected change over all maps with a single misplaced pixel by simply averaging over pixels:
\bea
\Delta^*_{-1}(f, G\mu) 
&\equiv&
\Bigg\langle
\log s(\xi^*_{-1}) - \log s(\xi^*)
\Bigg\rangle
 \\
&=& \frac{1}{N_{pix}} \sum_{i,j} \Big(  \log \big( s_{i,j}( 1-\xi^*_{i,j}) \big ) \; - \;   \log \big( s_{i,j}( \xi^*_{i,j})  \big ) \Big)
\nonumber
\eea
To recapitulate, $\Delta_{-1}(f, G\mu)$ is defined as the expected change in the log conditional probability of the evidence $f(\delta_{sky})$ as we condition on $\xi$ maps with only 1 pixel difference from the maximal map. Similarly, we define $\Delta_{-n}(f, G\mu)$ as the equivalent expectation taken over maps that differ by $n$ pixels from the maximal map. By independence of pixels we have (to a very good approximation when $n$ is low compared to the map size) that 
\[\Delta^*_{-n}(f, G\mu) = n \times \Delta^*_{-1}(f, G\mu)\]
i.e. the expected change in the log probability incured by changing $n$ pixels from the maximal map is simply $n$ times that of the expectation of changing $1$ pixel. This is not strictly correct, there are some errors involved from double counting the pixels, however, at low $n$ (where the probabilities are largest and most important) these errors are insignificant. If this error seems to become significant at high enough $n$, we cannot use the analytic method anymore and must resort to computing $\Delta^*_{-n}(f, G\mu)$ by drawing samples of $n$ pixels and observing the empirical expected change in log probability over multiple samples.

Using the equations above we can finally approximate $\log s( \xi)$ as 
\be
\label{approxlog}
\log s(\xi) \approx \log s(\xi^*) + n_{\xi^*, \xi} \times \Delta^*_{-1}(f, G\mu)
\ee
where $n_{\xi^*, \xi}$ is the number of differences between $\xi^*$ and $\xi$.

We use the results above to significantly simplify the posterior probability calculation currently given in terms of a sum over Boolean maps in equation~(\ref{PpGfsky}). 
Starting with equation~(\ref{PpGfsky}) and using equation~(\ref{approxlog}) we have
\begin{multline}
P'(G\mu \, | \,f(\delta_{sky})) \; 
=
P(G\mu) \; \sum_{\xi}  \frac{P( f(\delta_{sky})\, | \, \xi, G\mu)}{ P(\xi \, | \, f(\delta_{sky}), G\mu)} \times P(\xi)
 \\
~~~~~~~~~~~~~~~~~~~~~~~~~~~~= 
P(G\mu) \; \sum_{\xi} \exp \bigg ( \log s(\xi) \bigg ) \times P(\xi)
 \\
~~~~~~~\approx 
P(G\mu) \; \sum_{\xi} \exp \bigg ( \log s(\xi^*) + n_{\xi^*, \xi} \times \Delta^*_{-1}(f, G\mu)  \bigg ) \times P(\xi)
 \\
= 
P(G\mu) \; s(\xi^*) 
\sum_{\xi} \exp \bigg ( n_{\xi^*, \xi} \times \Delta^*_{-1}(f, G\mu) \bigg ) \times P(\xi)
\end{multline}
Notice that inside the sum over all Boolean maps $\xi$, maps contribute to the sum only through $ n_{\xi^*, \xi} $. Two maps for which this factor is identical will contribute the same amount to the sum (relative to their prior probabilities), hence we can write:
\begin{multline}
\sum_{\xi} \exp \bigg ( n_{\xi^*, \xi} \times \Delta^*_{-1}(f, G\mu) \bigg ) \times P(\xi) \\ =
\sum_{n} \exp \bigg ( n \times \Delta^*_{-1}(f, G\mu)\bigg ) \times \sum_{ n_{\xi^*, \xi}=n}  P(\xi) 
\end{multline}
All that remains to be done now is to compute the remaining sums over $\xi$ at the end of the last equation, we have:
\[\sum_{ n_{\xi^*, \xi}=n} P(\xi) = P(n | \xi^*)\]
i.e. the probability of a map $\xi$ having $n$ pixels different from $\xi^*$. 
We would like to calculate the probability $P(n | \xi^*)$ using histograms.  That is, we have the maps $\xi^*$ as well as a data set of maps $\xi^a$ sampled from $P(\xi)$, so we would like to simply directly evaluate $n_{\xi^*, \xi^a}$  and plot its histogram to get the probability distribution. However, the 450 map data set we used does not contains any maps at small $n$,
so we will not have any bins for those $n$ where the terms in the sum are largest.  For this reason we extrapolate the behaviour at low $n$ based on the behaviour at the $n$ we do encounter and we approximate $P(n | \xi^*)$ as a gaussian and compute the mean and standard deviation from data we do have. For the maps we use (see Section~\ref{nn}) the number of pixels is $N_{pix}=512\times512\approx 2.6 \times10^5$, and the mean and standard deviation of $P(n | \xi^*)$ are of order $10^5$ and $7000$, respectively.

To recap, we have transformed a sum with $2^{N_{pix}}$ terms (the sum over the Boolean maps with $N_{pix}$ pixels) into a sum over the integer $n$ between $0$ and $N_{pix}$. Needless to say, the latter is significantly more tractable than the former. Putting everything back together, we obtain that the posterior probability of $G\mu$ conditioned on the evidence produced by a function $f$ is
\begin{multline}
\label{Ppapprox_pre}
P'(G\mu \, | \,f(\delta_{sky})) \; \approx \; P(G\mu) \; \frac{P( f(\delta_{sky})\, | \, \xi^*, G\mu)}{ P(\xi^{*} \, | \, f(\delta_{sky}), G\mu)} \times 
\\
\sum_{n=0}^{N_{pix} } \exp \bigg ( n \times \Delta^*_{-1}(f, G\mu) ) \bigg) \times P(n | \xi^*)
\end{multline}
Using the gaussian calculation we present in Appendix~\ref{gauss_calc}, this can be rewritten  in terms of the mean $\mu$ and standard deviation $\sigma$ of $P(n | \xi^*)$ as
\begin{multline}
\label{Ppapprox}
P'(G\mu \, | \,f(\delta_{sky})) \; \approx \; 
\\
~~~~~~~~~~~~~~~P(G\mu) \; \frac{P( f(\delta_{sky})\, | \, \xi^*, G\mu)}{ P(\xi^{*} \, | \, f(\delta_{sky}), G\mu)} \times 
\\
~~~~~~~~~~~~~~~\exp \big ( \mu \Delta_{-1}^* +  \sigma^2 (\Delta_{-1}^*)^2)/2 \big ) \times 
\\
\frac{1}{2} \bigg (\text{Erfc} \bigg (  \frac{N_{pix}-(\mu + \sigma^2 \Delta_{-1}^*)}{\sqrt{2\sigma^2}} \bigg ) - \text{Erfc} \bigg ( - \frac{\mu + \sigma^2 \Delta_{-1}^*}{\sqrt{2\sigma^2}} \bigg ) \bigg )
\end{multline}
This is the formula we will use from now on to calculate the Bayesian posterior distribution of the string tension.

\section{A convolutional neural network as a choice for \lowercase{$\boldsymbol f$} }
\label{nn}

\subsection{The convolutional neural network of our previous work 
}
\label{oldnn}
One choice for the function $f$ is the five layer convolutional neural network described in detail in~\cite{Ciuca:2017ww}. The first layer involved a convolution map on the scalar valued pixels and gives a 32-dimensional value for each pixel.
Each subsequent layer of the network involves a convolution map on $N_{pix}$ elements with the following structure (the terms ''kernel size'' and ''stride'' are defined in~\cite{Ciuca:2017ww}):
%
\bea
{\rm layer\ 1:\ }  \text{~1-dim}&\to\text{32-dim},  {\rm kernel\ size=3, stride=1}
  \nonumber \\
    &\downarrow \tanh \nonumber \\
{\rm layer\ 2:\ }  \text{32-dim} &\to \text{32-dim},  {\rm kernel\ size=3, stride=1}
  \nonumber \\
    &\downarrow \tanh \nonumber \\
{\rm layer\ 3:\ }  \text{32-dim} &\to \text{32-dim},  {\rm kernel\ size=3, stride=1}
  \nonumber \\
    &\downarrow \tanh \nonumber \\
{\rm layer\ 4:\ }  \text{32-dim} &\to \text{32-dim},  {\rm kernel\ size=3, stride=1}
  \nonumber \\
    &\downarrow \tanh \nonumber \\
{\rm layer\ 5:\ }   \text{32-dim} &\to \text{1-dim},  {\rm kernel\ size=1, stride=1} 
\eea
From this we see that the network has $28\, 097$ parameters:
$$\big\{(32\cdot3^2+32)+(32^2\cdot3^2+32)\times3+32+1\big\}=28\, 097\ . $$ 

We trained this network by minimizing the cross entropy given by the Kullback-Leibler divergence between the probability $P(\xi \, | \, \delta_{sky}, G\mu)$ and the parametrized convolutional neural network that we used to approximate that probability [equation (4.7) in~\cite{Ciuca:2017jz} or equation (2.2) in~\cite{Ciuca:2017ww}].

\begin{figure}
\includegraphics[width=\columnwidth]{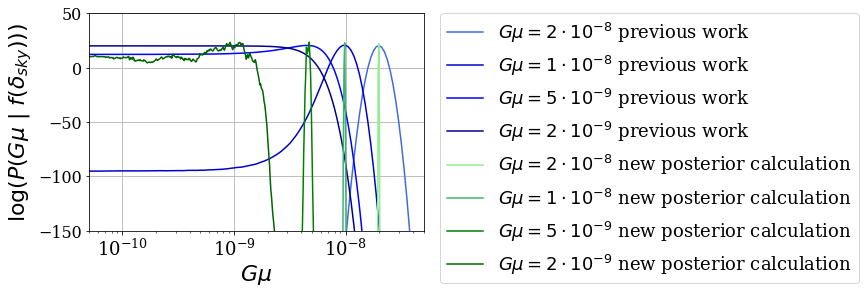}
\caption{Comparison of the posteriors obtained in our previous work  
and the posteriors calculated for the same network using equation~(\ref{Ppapprox}) .}
\label{fig:v2posteriorcompare}
\end{figure}

To obtain the posterior probability for the string tension in reference~\cite{Ciuca:2017jz}, we considered equations (2.4) or (4.8). Without the technique described in Section~\ref{more_efficient} above, this is a computationally difficult task so instead we approximated the posterior probability in the following way. We used the neural network $f$ to evaluate the sky map $\delta_{sky}$ with the unknown string tension and we binned the values $f_{i,j}(\delta_{sky})$ of each pixel into 1000 bins between 0 and 1. We then calculated the $\chi^2$ of the histogram $f(\delta_{sky})$ to a data set of histograms of $f_{i,j}$ values obtained from maps with known $G\mu$. This gave us an estimate of the string tension's posterior probability. In Fig.~\ref{fig:v2posteriorcompare} we compare this posterior probability from reference~\cite{Ciuca:2017jz} to that obtained by calculating with equation~(\ref{Ppapprox}). We see that the posterior estimates presented in~\cite{Ciuca:2017jz} were conservative and that the direct calculation of the Bayesian posterior probability of $G\mu$ gives much sharper peaks. 

\subsection{Residual networks: an improved convolutional neural network choice for ${\boldsymbol f}$}
\label{rn}
Simply adding more layers to the neural network in~\cite{Ciuca:2017jz} gave us a network that we were unable to train. By that we mean that the cross entropy did not decrease and converge to a lower value after each training iteration. One technique used to train deeper neural networks is to use a residual network~\cite{He:2015tt}. We experimented with residual networks between 5 and 100 layers and were able to have training converge for all of them. However we found that the results from using more than 30 layer network were not significantly better.  In this section we describe our residual network and then present and compare the results of the 30 layer residual network to those of our previous work~\cite{Ciuca:2017jz,Ciuca:2017ww}.

Whereas plain convolutional neural networks can only propagate forward to the next layer, residual networks allow for additional shortcut propagation from one layer to another one a few layers away.  These few layers that can be skipped over form what is called a residual block. 
Our 30 layer residual network consisted of 30 residual blocks sandwiched between an initial and a final layer. Each block contained three layers. 
Our residual network had the following structure for the convolution of each pixel:
\begin{gather}
{\rm initial\ layer:\ }\text{~1-dim}\to\text{32-dim},{\rm kernel\ size=3,}\ {\rm stride=1}
\nonumber\\
~~~~~~~~~~~~~~~~~~~~~~~~~~~~~~~\downarrow \nonumber\\
\text{\framebox{residual block 1: 32-dim$\to$32-dim}}
\nonumber\\
~~~~~~~~~~~~~~~~~~~~~~~~~~~~~~~\downarrow \nonumber\\
\text{\framebox{residual block 2: 32-dim$\to$32-dim}}
  \nonumber \\
~~~~~~~~~~~~~~~~~~~~~~~~~~~~~~~\downarrow \nonumber\\
~~~~~~~~~~~~~~~~~~~~~~~~~~~~~~~ ...  \\
~~~~~~~~~~~~~~~~~~~~~~~~~~~~~~~\downarrow \nonumber\\
\text{\framebox{residual block 30: 32-dim$\to$32-dim}}
  \nonumber \\
~~~~~~~~~~~~~~~~~~~~~~~~~~~~~~~\downarrow \nonumber\\
{\rm final\ layer\ 5:\ }\text{32-dim}\to \text{1-dim},  {\rm kernel\ size=1, stride=1} 
\nonumber
\end{gather}
In addition to being able to go through the block, one can go around the block and begin at the next block down. Each of the 30 residual blocks is composed of the following three layers:
\begin{align*}
{\rm layer\ 1:\ }   \text{32-dim} &\to \text{8-dim},  {\rm kernel\ size=1, stride=1} 
\\
&\downarrow \tanh  \\
{\rm layer\ 2:\ }   \text{~8-dim} &\to \text{8-dim},  {\rm kernel\ size=3, stride=1} 
\\%
&\downarrow \tanh  \\
{\rm layer\ 3:\ }   \text{~8-dim} &\to \text{32-dim},  {\rm kernel\ size=1, stride=1} 
\end{align*}
The number of parameters in our residual network is $65\, 857=$
$$\big\{(32+32)+(32^2+32)\times30+ (32\cdot8+8 +8^2\cdot3^2+8+8\cdot32+32)\times30  +  32+1\big\}. 
$$

We trained the residual network in the same way we trained our previous network~\citep{Ciuca:2017jz}. We used  numerically generated CMB temperature maps with and without cosmic strings. The data set was obtained with the same long string analytical model~\citep{Perivolaropoulos:1993efa} used in~\cite{Ciuca:2017jz} and other previous studies of cosmic string detection in CMB maps~\citep{Amsel:2008it,Stewart:2009fr,2010IJMPD..19..183D,Hergt:2017dr}. We used the PyTorch environment (pytorch.org) for machine learning and optimization algorithms, and we trained the model on a Tesla K80 GPU for 12 hours in total. 

\begin{figure}
\centering
\begin{subfigure}[b]{0.65\columnwidth}
\includegraphics[width=1.0\columnwidth]{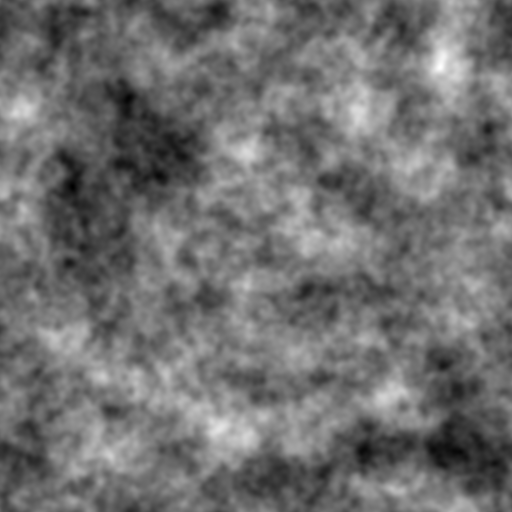}
\caption{The complete sky map, $\delta_{sky}$. Maps with and without strings are indistinguishable by eye.
}
\label{fig:skymap}
\end{subfigure}
\begin{subfigure}[b]{0.65\columnwidth}
\includegraphics[width=1.0\columnwidth]{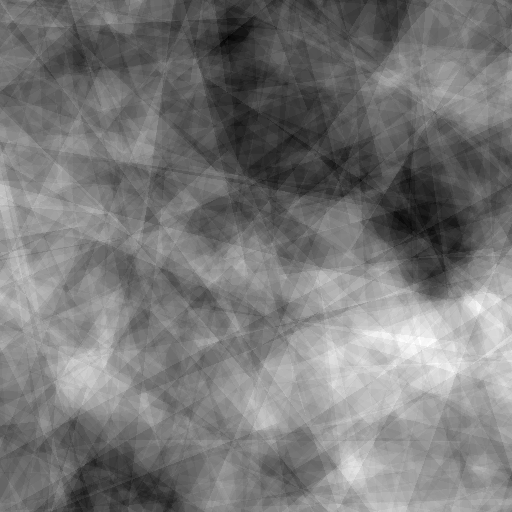}
\caption{String component contribution $\delta_{string}$ to the complete sky map.}
\label{fig:stringmap}
\end{subfigure}
\begin{subfigure}[b]{0.65\columnwidth}
\includegraphics[width=\columnwidth]{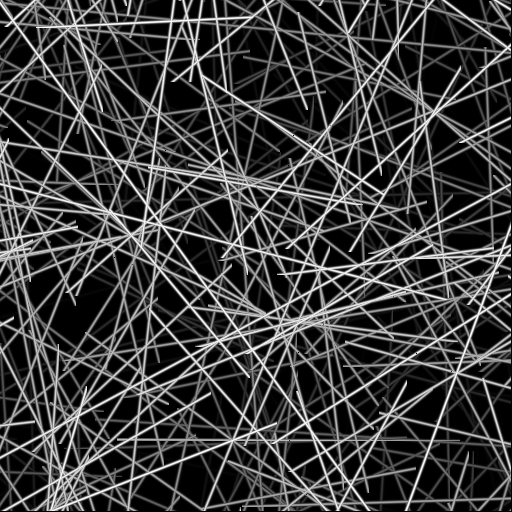}
\caption{String answer map $\xi$ used in the simulation.  }
\label{fig:strings}
\end{subfigure}
\caption{CMB anisotropy temperature maps of $512\times512$ pixels with a resolution of 1 arcmin per pixel. The white and black pixels are $+450\mu$K and $-450\mu$K anisotropies, respectively.  The shades of grey of the strings in the string answer map correspond to the relative strength of the string's GKS temperature discontinuity. 
(a) The complete sky map, $\delta_{sky}$. Maps with and without strings are indistinguishable by eye. (b) String component contribution $\delta_{string}$ to the complete sky map. (c) String answer map $\xi$ used in the simulation. 
}
\label{fig:maps}
\end{figure}

The maps were made up of $512\times512$ pixels with a resolution of 1 arcmin per pixel. We show these maps, which are the same as those in~\cite{Ciuca:2017jz}, in Fig.~\ref{fig:maps}.
The sky map is show in Fig.~\ref{fig:skymap}. For values of the string tension we study here, $G\mu\leq2\times 10^{-8}$, the sky map is indistinguishable by eye from a pure Gaussian fluctuation map (i.e. $G\mu=0$). 
The string temperature component to the complete sky map is shown in Fig.~\ref{fig:stringmap} with a $G\mu=1$. The string answer map is shown in Fig.~\ref{fig:strings}. 
One of the unknown parameters characterizing the scaling solution of strings 
is the number of strings per Hubble volume, $N_H$, which can have a value between 1 and 10. We trained our neural network with a value of $N_H=1$ and this did not impair the predictive power for input maps with larger $N_H$ values. This is an indication that the network is indeed generalizing and not just overfitting.

\subsection{Residual network prediction maps and ${\pmb G\mu}$ posteriors}
\label{rnpredictions}

\begin{figure*}
\centering
\begin{subfigure}[b]{0.6\columnwidth}
\includegraphics[width=\columnwidth]{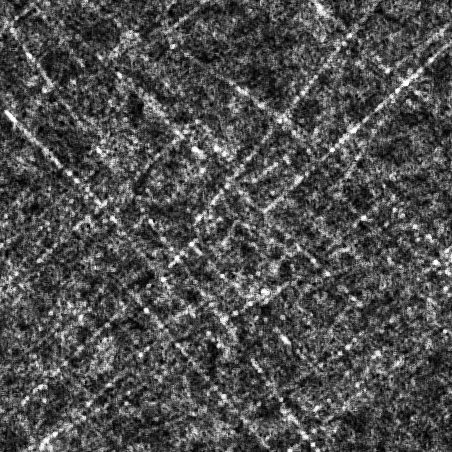}
\caption{Reference \protect{\cite{Ciuca:2017jz}} prediction $G\mu=10^{-8}$}
\label{prediction_v2_1e-8}
\end{subfigure}
\begin{subfigure}[b]{0.6\columnwidth}
\includegraphics[width=\columnwidth]{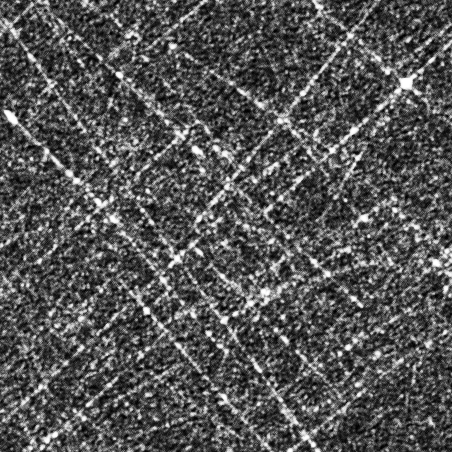}
\caption{Residual network prediction \\ $ G\mu=10^{-8}$}
\label{prediction_v6_1e-8}
\end{subfigure}
\\ 
\begin{subfigure}[b]{0.6\columnwidth}
\includegraphics[width=\columnwidth]{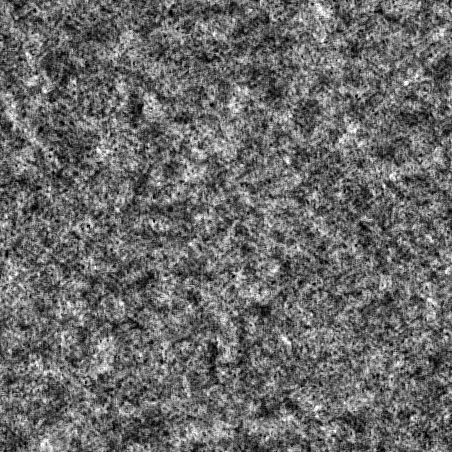}
\caption{Reference \protect{\cite{Ciuca:2017jz}} prediction $G\mu=5\times10^{-9}$}
\label{prediction_v2_5e-9}
\end{subfigure}
\begin{subfigure}[b]{0.6\columnwidth}
\includegraphics[width=\columnwidth]{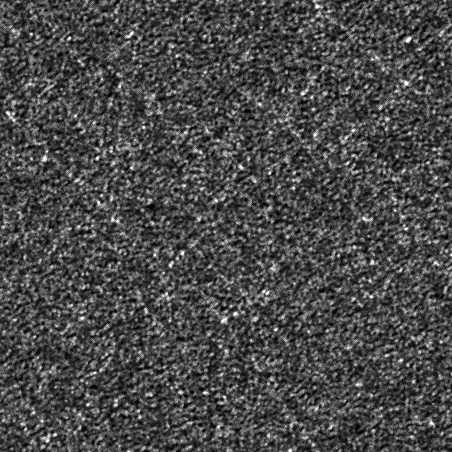}
\caption{Residual network prediction \\ $G\mu=5\times10^{-9}$}
\label{prediction_v6_5e-9}
\end{subfigure}
\\
\begin{subfigure}[b]{0.6\columnwidth}
\includegraphics[width=\columnwidth]{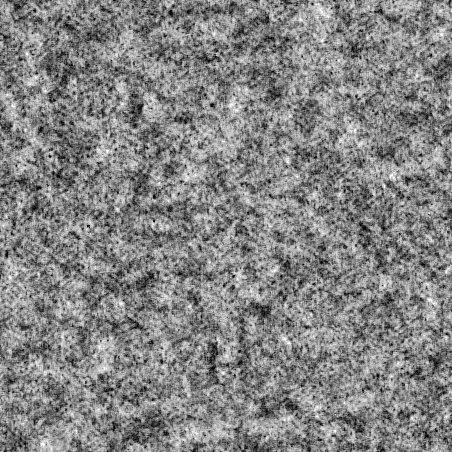}
\caption{Reference \protect{\cite{Ciuca:2017jz}} prediction $G\mu=2\times10^{-9}$}
\label{prediction_v2_2e-9}
\end{subfigure}
\begin{subfigure}[b]{0.6\columnwidth}
\includegraphics[width=\columnwidth]{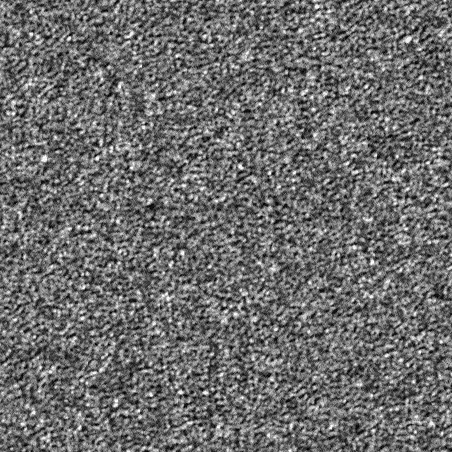}
\caption{Residual network prediction \\ $G\mu=2\times10^{-9}$}
\label{prediction_v6_2e-9}
\end{subfigure}
\caption{Comparison of neural network predictions without noise.
The actual placement of long strings, the $\xi$ map, is given in Fig.~\ref{fig:strings}. We compare our neural network's prediction of $\xi$ for different value of the string tension with no noise for the network used in our previous work,
on the left, and the residual network introduced here on the right. The shades of grey in the prediction maps correspond to the probability of a pixel being on a string, with completely black pixels being 0 probability and completely white pixels being probability 1. All the figures correspond to $512\times512$ pixels with a resolution of 1 arcmin per pixel. }
\label{stringpredictions}
\end{figure*}

In Fig.~\ref{stringpredictions} we show our residual and plain network predictions for the string location map 
using different values for $G\mu$, with $N_H=3$, and no noise. 
The shades of grey in the prediction maps correspond to the probability of a pixel being on a string. Completely black pixels are probability 0 and completely white pixels are probability 1 of being on a string.  As $G\mu$ tends to zero, the neural network provides less information as to whether a pixel is on a string or not and the pixel probabilities tend to the prior $P((i,j)\in string)$ which is given by the number of pixels on strings in the answer map $\xi$ (fig.~\ref{fig:strings}) divided by the total number of pixels. Thus as $G\mu$ tends to zero, our prediction map will become more uniformly grey, as figures \ref{prediction_v2_2e-9} and \ref{prediction_v6_2e-9} show. 

\begin{figure}
\includegraphics[width=\columnwidth]{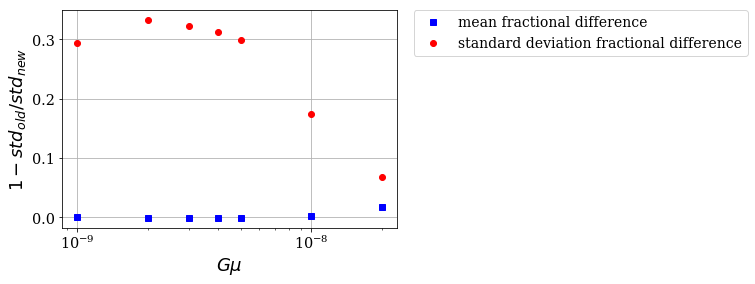}
\caption{Comparison of mean and standard deviation over pixels in the prediction maps of the two networks. We plot the fractional difference between the two networks, $1-\frac{\rm old\ network}{\rm residual\ network}$.
}
\label{fig:meanstd}
\end{figure}

The two predictions maps look different, though it is not immediately clear that the residual network is better.  However with a  careful visual comparison of the prediction maps from the two networks we see that the residual network is distinguishing the string locations more clearly. In fact, we show in Fig.~\ref{fig:meanstd} that the prediction values in the residual network have a standard deviation that is larger than our old network. The mean value of the predictions is the same, yet the residual network assigns more high and low probability values than our old network. In other words, the residual network's certainty of which pixels contain strings, and which pixels do not, is greater than for that of the old network.

\begin{figure}
\includegraphics[width=\columnwidth]{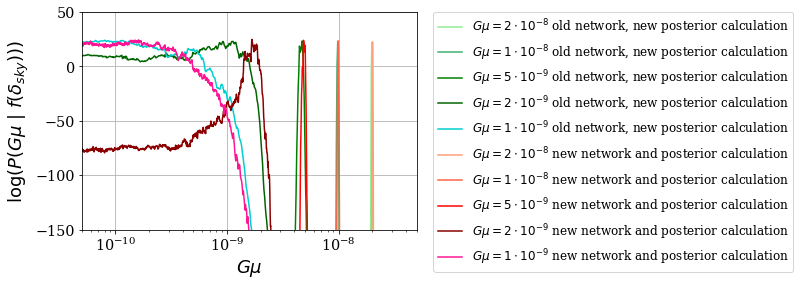}
\caption{Comparison of the posteriors' improvement obtained by using the residual network in place of the plain convolutional neural network. 
In both cases the posteriors are calculated using equation~(\ref{Ppapprox}).
}
\label{fig:nn_improvement}
\end{figure}

However, the real test of the superior performance of the residual network is the posterior probability of the string tension that it provides through equation~(\ref{Ppapprox}). 
In Fig.~\ref{fig:nn_improvement} we compare the posteriors obtained from the old neural network to those obtained from the residual network, both calculated with equation~(\ref{Ppapprox}). The improvement shown in Fig.~\ref{fig:nn_improvement} plus the improvement presented in Fig.~\ref{fig:v2posteriorcompare} results in the total improvement 
we presented in Fig.~\ref{fig:v2paper_to_v6resnet} of the introduction.

From Fig.~\ref{fig:nn_improvement} we see that all the peaks in the posterior distribution from the residual network are sharper and more accurately centred over the true value of $G\mu$. For $G\mu=2\times10^{-9}$ the residual network provides a clear bump over the correct string tension, whereas the old network does not. Calculating the area under the probability given by this bump tells us that there is a 0.99 probability that $G\mu\in[1.7,2.1]\times10^{-9}$.

\begin{figure}
\includegraphics[width=\columnwidth]{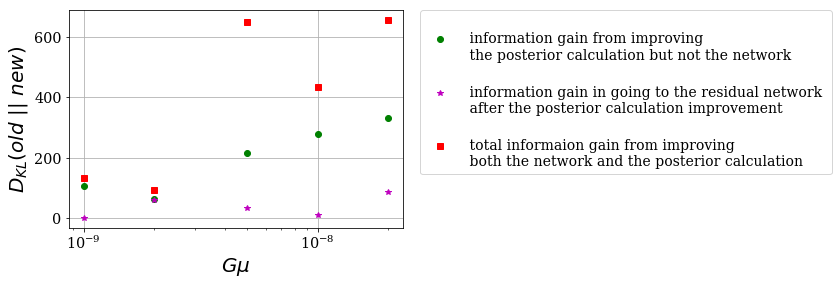}
\caption{Information gain on the posterior distribution of the string tension from analysing sky maps with the residual network and the new calculation of the posterior using equation~(\ref{Ppapprox}), versus the plain convolutional neural network and the calculation from our previous work. 
 }
\label{fig:DKL}
\end{figure}

The Kullback-Leibler divergence allows us to quantify how much more information is gained in going from the posterior probabilities of our old network $P_{old}$, to that of the residual network $P_{new}$: 
\bea\label{KLdiv}
D_{KL}(P_{old} || P_{new}) \equiv & \nonumber\\
& \int d(G\mu) ~ P_{old}(G\mu) \log\frac{P_{old}(G\mu)}{P_{new}(G\mu)} 
\eea
In Fig.~\ref{fig:DKL} we plot the value of the KL divergence at various string tensions between the two posteriors given in Fig.s~\ref{fig:v2paper_to_v6resnet},\ref{fig:v2posteriorcompare},\ref{fig:nn_improvement}. 


\section{Conclusions and a new optimization goal for finding \lowercase{$\boldsymbol f$}}
\label{conclusions}
In our previous work~\citep{Ciuca:2017jz,Ciuca:2017ww} we presented a Bayesian interpretation of cosmic string detection in which the posterior probability distribution of the string tension is linked to the estimates of cosmic string location on a CMB map. Here, we have presented a reformulation of the posterior formula and introduced a more efficient and accurate way to compute the posterior probability. In addition we have improved our convolutional neural network with residual networks to yield better prediction maps and posterior probabilities for the string tension. We have presented and quantified these improvements in subsection~\ref{rn} and~\ref{rnpredictions}. All these improvements together can be summarized by Fig.~\ref{fig:v2paper_to_v6resnet} in the Introduction section. 

We have improved our previous neural network by using the more sophisticated architecture provided by residual networks~\cite{He:2015tt} as a preamble to introducing noise and then applying our analysis to realistic string simulations~(\cite{Fraisse:2007nu,PlanckCollaboration:2014il}; work in progress). While we have presented our analysis in the context of cosmic string detection in CMB temperature anisotropy maps, generalizing the procedure to cosmic string wake detection in 21 cm intensity maps is straightforward~\citep{Brandenberger:2010hi, Hernandez:2011ima, Hernandez:2012gz, Hernandez:2014cu, daCunha:2016bo}.

Our derivation of a posterior probability for the string tension was specifically done in order to derive a connection between the string tension and the location of cosmic strings on a map. We then trained our neural network $f$ to be good at estimating string locations. We did this by minimizing the KL divergence between 
$f(\delta_{sky})$
and $P(\xi | \delta_{sky})$. 
Thus "good" functions $f$ are those that produce outputs that resemble the true answer map $\xi$. 
The network has no knowledge of our Bayesian formula connecting these string locations to the string tension.  And while we can experimentally verify that the function that optimizes the KL divergence leads to good posteriors, it is unlikely to be the best one.
All this suggests that another way of obtaining $f$ would be to directly optimize the posterior probability distributions. Since the derivations of section~\ref{better_posterior} were all made in a way agnostic to $f$, any function $f: R^{N_{pix}} \rightarrow R^{N_{pix}}$ can be plugged-in the calculations and the procedure will spit out the corresponding posteriors. 
Hence, the Bayesian procedure derived in the previous sections implicitly provides a criterion to evaluate the "goodness" of various functions $f$ : on average, does $f$ lead to sharp posteriors centred around the correct $G\mu$? It would be interesting to see if we still obtain "good" prediction maps when $f$ is chosen in this way. 

\section*{Acknowledgements}
We thank Christophe Ringeval for useful discussions.
We acknowledge the support of the Fonds de recherche du Qu\'ebec -- Nature et technologies (FRQNT) Programme de recherche pour les enseignants de coll\`ege, 
and the support of the Natural Sciences and Engineering Research Council of Canada (NSERC) (funding reference number SAPIN-2018-00020).
Computations were made on the supercomputer Helios from Universit\'e Laval, managed by Calcul Qu\'ebec and Compute Canada. The operation of this supercomputer is funded by the Canada Foundation for Innovation (CFI), the minist\`ere de l'\'Economie, de la science et de l'innovation du Qu\'ebec (MESI) and the Fonds de recherche du Qu\'ebec -- Nature et technologies (FRQNT). 

\bibliographystyle{mnras}
\bibliography{Neural_Net_Cosmic_Strings_III_v2.5}

\begin{thebibliography}{}
\makeatletter
\relax
\def\mn@urlcharsother{\let\do\@makeother \do\$\do\&\do\#\do\^\do\_\do\%\do\~}
\def\mn@doi{\begingroup\mn@urlcharsother \@ifnextchar [ {\mn@doi@}
  {\mn@doi@[]}}
\def\mn@doi@[#1]#2{\def\@tempa{#1}\ifx\@tempa\@empty \href
  {http://dx.doi.org/#2} {doi:#2}\else \href {http://dx.doi.org/#2} {#1}\fi
  \endgroup}
\def\mn@eprint#1#2{\mn@eprint@#1:#2::\@nil}
\def\mn@eprint@arXiv#1{\href {http://arxiv.org/abs/#1} {{\tt arXiv:#1}}}
\def\mn@eprint@dblp#1{\href {http://dblp.uni-trier.de/rec/bibtex/#1.xml}
  {dblp:#1}}
\def\mn@eprint@#1:#2:#3:#4\@nil{\def\@tempa {#1}\def\@tempb {#2}\def\@tempc
  {#3}\ifx \@tempc \@empty \let \@tempc \@tempb \let \@tempb \@tempa \fi \ifx
  \@tempb \@empty \def\@tempb {arXiv}\fi \@ifundefined
  {mn@eprint@\@tempb}{\@tempb:\@tempc}{\expandafter \expandafter \csname
  mn@eprint@\@tempb\endcsname \expandafter{\@tempc}}}

\bibitem[\protect\citeauthoryear{Amsel, Berger  \& Brandenberger}{Amsel
  et~al.}{2008}]{Amsel:2008it}
Amsel S.,  Berger J.,   Brandenberger R.~H.,  2008, \mn@doi [Journal of
  Cosmology and Astroparticle Physics] {10.1088/1475-7516/2008/04/015}, 2008,
  015

\bibitem[\protect\citeauthoryear{Blanco-Pillado, Olum  \&
  Shlaer}{Blanco-Pillado et~al.}{2014}]{BlancoPillado:2014kr}
Blanco-Pillado J.~J.,  Olum K.~D.,   Shlaer B.,  2014, \mn@doi [Physical Review
  D] {10.1103/PhysRevD.89.023512}, 89, 023512

\bibitem[\protect\citeauthoryear{Brandenberger, Danos, Hern{\'a}ndez  \&
  Holder}{Brandenberger et~al.}{2010}]{Brandenberger:2010hi}
Brandenberger R.~H.,  Danos R.~J.,  Hern{\'a}ndez O.~F.,   Holder G.~P.,  2010,
  \mn@doi [Journal of Cosmology and Astroparticle Physics]
  {10.1088/1475-7516/2010/12/028}, 2010, 028

\bibitem[\protect\citeauthoryear{Canny}{Canny}{1986}]{Canny:et}
Canny J.,  1986, \mn@doi [IEEE Transactions on Pattern Analysis and Machine
  Intelligence] {10.1109/TPAMI.1986.4767851}, PAMI-8, 679

\bibitem[\protect\citeauthoryear{Chown et~al.,}{Chown
  et~al.}{2018}]{Chown:2018ez}
Chown R.,  et~al., 2018, \mn@doi [The Astrophysical Journal Supplement Series]
  {10.3847/1538-4365/aae694}, 239, 10

\bibitem[\protect\citeauthoryear{Ciuca \& Hern{\'a}ndez}{Ciuca \&
  Hern{\'a}ndez}{2017}]{Ciuca:2017jz}
Ciuca R.,  Hern{\'a}ndez O.~F.,  2017, \mn@doi [Journal of Cosmology and
  Astroparticle Physics] {10.1088/1475-7516/2017/08/028}, 2017, 028

\bibitem[\protect\citeauthoryear{Ciuca, Hern{\'a}ndez  \& Wolman}{Ciuca
  et~al.}{2017}]{Ciuca:2017ww}
Ciuca R.,  Hern{\'a}ndez O.~F.,   Wolman M.,  2017, arXiv.org, p.
  arXiv:1708.08878

\bibitem[\protect\citeauthoryear{Danos \& Brandenberger}{Danos \&
  Brandenberger}{2010}]{2010IJMPD..19..183D}
Danos R.~J.,  Brandenberger R.~H.,  2010, \mn@doi [International Journal of
  Modern Physics D] {10.1142/S0218271810016324}, 19, 183

\bibitem[\protect\citeauthoryear{Fraisse, Ringeval, Spergel  \&
  Bouchet}{Fraisse et~al.}{2008}]{Fraisse:2007nu}
Fraisse A.~A.,  Ringeval C.,  Spergel D.~N.,   Bouchet F.~R.,  2008, \mn@doi
  [Physical Review D] {10.1103/PhysRevD.78.043535}, D78, 043535

\bibitem[\protect\citeauthoryear{Gott}{Gott}{1985}]{Gott:1985eg}
Gott J. R.~I.,  1985, \mn@doi [The Astrophysical Journal] {10.1086/162808},
  288, 422

\bibitem[\protect\citeauthoryear{He, Zhang, Ren  \& Sun}{He
  et~al.}{2015}]{He:2015tt}
He K.,  Zhang X.,  Ren S.,   Sun J.,  2015, arXiv.org, p. arXiv:1512.03385

\bibitem[\protect\citeauthoryear{Hergt, Amara, Brandenberger, Kacprzak  \&
  Refregier}{Hergt et~al.}{2017}]{Hergt:2017dr}
Hergt L.,  Amara A.,  Brandenberger R.~H.,  Kacprzak T.,   Refregier A.,  2017,
  \mn@doi [Journal of Cosmology and Astroparticle Physics]
  {10.1088/1475-7516/2017/06/004}, 2017, 004

\bibitem[\protect\citeauthoryear{Hern{\'a}ndez}{Hern{\'a}ndez}{2014}]{Hernandez:2014cu}
Hern{\'a}ndez O.~F.,  2014, \mn@doi [Physical Review D]
  {10.1103/PhysRevD.90.123504}, 90, 123504

\bibitem[\protect\citeauthoryear{Hern{\'a}ndez \& Brandenberger}{Hern{\'a}ndez
  \& Brandenberger}{2012}]{Hernandez:2012gz}
Hern{\'a}ndez O.~F.,  Brandenberger R.~H.,  2012, \mn@doi [Journal of Cosmology
  and Astroparticle Physics] {10.1088/1475-7516/2012/07/032}, 2012, 032

\bibitem[\protect\citeauthoryear{Hern{\'a}ndez, Wang, Fong  \&
  Brandenberger}{Hern{\'a}ndez et~al.}{2011}]{Hernandez:2011ima}
Hern{\'a}ndez O.~F.,  Wang Y.,  Fong J.,   Brandenberger R.~H.,  2011, \mn@doi
  [Journal of Cosmology and Astroparticle Physics]
  {10.1088/1475-7516/2011/08/014}, 2011, 014

\bibitem[\protect\citeauthoryear{Hindmarsh, Lizarraga, Urrestilla, Daverio  \&
  Kunz}{Hindmarsh et~al.}{2017}]{Hindmarsh:2017iw}
Hindmarsh M.,  Lizarraga J.,  Urrestilla J.,  Daverio D.,   Kunz M.,  2017,
  \mn@doi [Physical Review D] {10.1103/PhysRevD.96.023525}, 96, 023525

\bibitem[\protect\citeauthoryear{Jeong \& Smoot}{Jeong \&
  Smoot}{2005}]{Jeong:2005bg}
Jeong E.,  Smoot G.~F.,  2005, \mn@doi [The Astrophysical Journal]
  {10.1086/428921}, 624, 21

\bibitem[\protect\citeauthoryear{Jeong \& Smoot}{Jeong \&
  Smoot}{2007}]{Jeong:2007jl}
Jeong E.,  Smoot G.~F.,  2007, \mn@doi [The Astrophysical Journal]
  {10.1086/518556}, 661, L1

\bibitem[\protect\citeauthoryear{Jeong, Baccigalupi  \& Smoot}{Jeong
  et~al.}{2010}]{Jeong:2010dh}
Jeong E.,  Baccigalupi C.,   Smoot G.~F.,  2010, \mn@doi [Journal of Cosmology
  and Astroparticle Physics] {10.1088/1475-7516/2010/09/018}, 2010, 018

\bibitem[\protect\citeauthoryear{Kaiser \& Stebbins}{Kaiser \&
  Stebbins}{1984}]{Kaiser:1984jg}
Kaiser N.,  Stebbins A.,  1984, \mn@doi [Nature] {10.1038/310391a0}, 310, 391

\bibitem[\protect\citeauthoryear{Lo \& Wright}{Lo \& Wright}{2005}]{Lo:2005tm}
Lo A.~S.,  Wright E.~L.,  2005, arXiv.org, pp arXiv:astro--ph--0503120

\bibitem[\protect\citeauthoryear{Lorenz, Ringeval  \& Sakellariadou}{Lorenz
  et~al.}{2010}]{Lorenz:2010iq}
Lorenz L.,  Ringeval C.,   Sakellariadou M.,  2010, \mn@doi [Journal of
  Cosmology and Astroparticle Physics] {10.1088/1475-7516/2010/10/003}, 2010,
  003

\bibitem[\protect\citeauthoryear{Louis et~al.,}{Louis
  et~al.}{2014}]{Louis:2014ef}
Louis T.,  et~al., 2014, \mn@doi [Journal of Cosmology and Astroparticle
  Physics] {10.1088/1475-7516/2014/07/016}, 2014, 016

\bibitem[\protect\citeauthoryear{McEwen, Feeney, Peiris, Wiaux, Ringeval  \&
  Bouchet}{McEwen et~al.}{2017}]{McEwen:2017cg}
McEwen J.~D.,  Feeney S.~M.,  Peiris H.~V.,  Wiaux Y.,  Ringeval C.,   Bouchet
  F.~R.,  2017, \mn@doi [Monthly Notices of the Royal Astronomical Society]
  {10.1093/mnras/stx2268}, 472, 4081

\bibitem[\protect\citeauthoryear{Movahed \& Khosravi}{Movahed \&
  Khosravi}{2011}]{Movahed:2011em}
Movahed M.~S.,  Khosravi S.,  2011, \mn@doi [Journal of Cosmology and
  Astroparticle Physics] {10.1088/1475-7516/2011/03/012}, 2011, 012

\bibitem[\protect\citeauthoryear{Perivolaropoulos}{Perivolaropoulos}{1993}]{Perivolaropoulos:1993efa}
Perivolaropoulos L.,  1993, \mn@doi [Physics Letters B]
  {10.1016/0370-2693(93)91825-8}, 298, 305

\bibitem[\protect\citeauthoryear{{Planck Collaboration} et~al.,}{{Planck
  Collaboration} et~al.}{2014}]{PlanckCollaboration:2014il}
{Planck Collaboration} et~al., 2014, \mn@doi [Astronomy and Astrophysics]
  {10.1051/0004-6361/201321621}, 571, A25

\bibitem[\protect\citeauthoryear{Ringeval \& Bouchet}{Ringeval \&
  Bouchet}{2012}]{Ringeval:2012gp}
Ringeval C.,  Bouchet F.~R.,  2012, \mn@doi [Physical Review D]
  {10.1103/PhysRevD.86.023513}, 86, 16464

\bibitem[\protect\citeauthoryear{Ringeval, Sakellariadou  \& Bouchet}{Ringeval
  et~al.}{2007}]{Ringeval:2007gf}
Ringeval C.,  Sakellariadou M.,   Bouchet F.~R.,  2007, \mn@doi [Journal of
  Cosmology and Astroparticle Physics] {10.1088/1475-7516/2007/02/023}, 2007,
  023

\bibitem[\protect\citeauthoryear{Stewart \& Brandenberger}{Stewart \&
  Brandenberger}{2009}]{Stewart:2009fr}
Stewart A.,  Brandenberger R.~H.,  2009, \mn@doi [Journal of Cosmology and
  Astroparticle Physics] {10.1088/1475-7516/2009/02/009}, 2009, 009

\bibitem[\protect\citeauthoryear{Vafaei~Sadr, Movahed, Farhang, Ringeval,
  Bouchet  \& Bouchet}{Vafaei~Sadr et~al.}{2018a}]{VafaeiSadr:2018hh}
Vafaei~Sadr A.,  Movahed S. M.~S.,  Farhang M.,  Ringeval C.,  Bouchet F.~R.,
  Bouchet F.~R.,  2018a, \mn@doi [Monthly Notices of the Royal Astronomical
  Society] {10.1093/mnras/stx3126}, 475, 1010

\bibitem[\protect\citeauthoryear{Vafaei~Sadr, Farhang, Movahed, Bassett  \&
  Kunz}{Vafaei~Sadr et~al.}{2018b}]{VafaeiSadr:2018bc}
Vafaei~Sadr A.,  Farhang M.,  Movahed S. M.~S.,  Bassett B.,   Kunz M.,  2018b,
  \mn@doi [Monthly Notices of the Royal Astronomical Society]
  {10.1093/mnras/sty1055}, 478, 1132

\bibitem[\protect\citeauthoryear{da Cunha, Brandenberger  \&
  Hern{\'a}ndez}{da~Cunha et~al.}{2016}]{daCunha:2016bo}
da Cunha D. C.~N.,  Brandenberger R.~H.,   Hern{\'a}ndez O.~F.,  2016, \mn@doi
  [Physical Review D] {10.1103/PhysRevD.93.123501}, 93, 123501

\makeatother
\end{thebibliography}

\appendix
\section{Derivation of equation~(2) from Bayes theorem} 
\label{bayes_derivation}
Consider:
\[ P(\xi \, | \, f(\delta_{sky}), G\mu) \; = \; \frac{P( f(\delta_{sky}), G\mu \, | \, \xi ) P(\xi)}{ P(f(\delta_{sky}), G\mu)} \]

\[ P(\xi \, | \, f(\delta_{sky}), G\mu) \; = \; \frac{P( f(\delta_{sky}), G\mu \, | \, \xi ) P(\xi)}{ P(G\mu \, | \,f(\delta_{sky})) P(f(\delta_{sky}))} \]
shufling terms around and using $P( f(\delta_{sky}), G\mu \, | \, \xi ) \, = \, P( f(\delta_{sky})\, | \, \xi, G\mu) P(G\mu \, | \, \xi) \, = \, P( f(\delta_{sky})\, | \, \xi, G\mu) P(G\mu) $:

\[ P(G\mu \, | \,f(\delta_{sky})) \; = \; \frac{P( f(\delta_{sky})\, | \, \xi, G\mu) \times P(G\mu) \times  P(\xi)}{ P(\xi \, | \, f(\delta_{sky}), G\mu) \times P(f(\delta_{sky}))} \]

Summing over all Boolean maps $\xi$:

\[ 2^{N_{pix}} P(G\mu \, | \,f(\delta_{sky})) \; = \; \sum_{\xi} \frac{P( f(\delta_{sky})\, | \, \xi, G\mu) \times P(G\mu) \times  P(\xi)}{ P(\xi \, | \, f(\delta_{sky}), G\mu) \times P(f(\delta_{sky}))}\]

\[ 2^{N_{pix}} P(G\mu \, | \,f(\delta_{sky})) \; = \frac{P(G\mu)}{P(f(\delta_{sky}))} \; \sum_{\xi} \frac{P( f(\delta_{sky})\, | \, \xi, G\mu)  \times  P(\xi)}{ P(\xi \, | \, f(\delta_{sky}), G\mu)} \]

Transforming the sum into an expectation over maps $\xi^a$ sampled from $P(\xi)$ and switching the factor of $2^{N_{pix}}$ to the right:

\[ P(G\mu \, | \,f(\delta_{sky})) \; = 2^{-N_{pix}}\frac{P(G\mu)}{P(f(\delta_{sky}))} \; \frac{1}{N}\sum_{\xi^a \sim P(\xi)} \frac{P( f(\delta_{sky})\, | \, \xi^a, G\mu)}{ P(\xi^a \, | \, f(\delta_{sky}), G\mu)} \]

\section{Small gaussian calculation}
\label{gauss_calc}

We present the calculation used to go from equation~(\ref{Ppapprox_pre}) to~(\ref{Ppapprox}).

\begin{multline}\sum_{n=0}^{N} \exp \big ( n\times \Delta^*_{-1} \big) \frac{1}{\sqrt{2\pi \sigma^2}} \exp \big ( -\frac{(n-\mu)^2}{2\sigma^2} \big ) \nonumber\end{multline}

\begin{multline}=\sum_{n=0}^{N} \frac{1}{\sqrt{2\pi \sigma^2}} \exp \big ( n\Delta^*_{-1} -\frac{(n-\mu)^2}{2\sigma^2} \big ) \nonumber\end{multline}

\begin{multline}=\sum_{n=0}^{N} \frac{1}{\sqrt{2\pi \sigma^2}} \exp \big (  -\frac{(n-\mu)^2 - 2\sigma^2 n\Delta^*_{-1}}{2\sigma^2} \big ) \nonumber\end{multline}

\begin{multline}=\sum_{n=0}^{N} \frac{1}{\sqrt{2\pi \sigma^2}} \exp \big (  -\frac{n^2 - 2n\mu + \mu^2  - 2\sigma^2 n\Delta^*_{-1}}{2\sigma^2} \big ) \nonumber\end{multline}

\begin{multline}=\sum_{n=0}^{N} \frac{1}{\sqrt{2\pi \sigma^2}} \exp \big (  -\frac{(n - (\mu + \sigma^2 \Delta_{-1}^*))^2  + \mu^2 - (\mu + \sigma^2 \Delta_{-1}^*)^2}{2\sigma^2} \big ) \nonumber\end{multline}

\begin{multline}=\exp \big ( - \frac{\mu^2 - (\mu + \sigma^2 \Delta_{-1}^*)^2}{2\sigma^2} \big ) \times \int_{0}^{N} \frac{1}{\sqrt{2\pi \sigma^2}} \exp \big (  -\frac{(n - (\mu + \sigma^2 \Delta_{-1}^*))^2}{2\sigma^2}\big)\nonumber\end{multline}

\begin{multline}=\exp \big ( - \frac{ - 2\mu \sigma^2 \Delta_{-1}^* - ( \sigma^2 \Delta_{-1}^*)^2)}{2\sigma^2} \big ) 
\\ \times \frac{1}{2} \bigg ( -\text{Erf} \bigg ( \frac{(\mu + \sigma^2 \Delta_{-1}^*)-N}{\sqrt{2\sigma^2}} \bigg ) + \text{Erf} \bigg ( \frac{\mu + \sigma^2 \Delta_{-1}^*}{\sqrt{2\sigma^2}} \bigg ) \bigg )\nonumber\end{multline}

\begin{multline}=\exp \big ( \mu \Delta_{-1}^* +  \sigma^2 (\Delta_{-1}^*)^2)/2 \big ) 
\\\times \frac{1}{2} \bigg ( -\text{Erf} \bigg ( \frac{(\mu + \sigma^2 \Delta_{-1}^*)-N}{\sqrt{2\sigma^2}} \bigg ) + \text{Erf} \bigg ( \frac{\mu + \sigma^2 \Delta_{-1}^*}{\sqrt{2\sigma^2}} \bigg ) \bigg )\nonumber\end{multline}

\begin{multline}
=\exp \big ( \mu \Delta_{-1}^* +  \sigma^2 (\Delta_{-1}^*)^2)/2 \big ) 
\\ \times \frac{1}{2} \bigg (\text{Erfc} \bigg (  \frac{N-(\mu + \sigma^2 \Delta_{-1}^*)}{\sqrt{2\sigma^2}} \bigg ) - \text{Erfc} \bigg ( - \frac{\mu + \sigma^2 \Delta_{-1}^*}{\sqrt{2\sigma^2}} \bigg ) \bigg )
\nonumber\end{multline}

\bsp	
\label{lastpage}
\end{document}